\documentclass[doublecol]{epl2} 

\usepackage{latexsym,amsmath,amssymb}
\usepackage{dcolumn}
\usepackage{bm}
\usepackage[usenames,dvipsnames]{xcolor}
\usepackage{epstopdf}
\usepackage[a4paper,colorlinks=true,linkcolor=Blue,citecolor=Blue,urlcolor=Blue]{hyperref}
\usepackage{xspace}

\newcommand{\bra}[1]{\langle #1|}
\newcommand{\ket}[1]{|#1\rangle}

\newcommand{\e}{\varepsilon}
\newcommand{\s}{\sigma}
\newcommand{\G}{\Gamma}
\newcommand{\be}{\begin{equation}}
\newcommand{\ee}{\end{equation}}
\newcommand{\beq}{\begin{eqnarray}}
\newcommand{\eeq}{\end{eqnarray}}

\newcommand{\fig}[1]{Fig.~\ref{#1}}
\newcommand{\Q}[1]{\color{red}}

\title{Tunnel magnetoresistance of a supramolecular spin valve}

\author{Anna P\l{}omi\'nska \and Ireneusz Weymann}
\shortauthor{A. P\l{}omi\'nska and I. Weymann}

\institute{
  Faculty of Physics, Adam Mickiewicz University, Pozna\'{n}, Poland
}


\abstract{
We theoretically study the transport properties of a supramolecular spin valve,
consisting of a carbon nanotube with two attached magnetic molecules,
weakly coupled to metallic contacts.
The emphasis is put on analyzing the change of the system's transport properties
with the application of an external magnetic field,
which aligns the spins of the molecules.
It is shown that magnetoresistive properties of the considered molecular junction,
which are associated with changing the state of the molecules
from superparamagnetic to the ferromagnetic one,
strongly depend on the applied bias voltage and the position of the nanotube's orbital
levels, which can be tuned by a gate voltage.
A strong dependence on the transport regime
is also found in the case of the spin polarization of the current
flowing through the system. The mechanisms leading to those effects are
explained by invoking appropriate molecular states responsible for transport.
The analysis is done with aid of the real-time diagrammatic technique
up to the second order of expansion with respect to tunneling processes.
}

\begin{document}

\maketitle

\section{Introduction}

Over the past decades the transport properties of nano-scale devices based
on magnetic molecules have been intensively studied
\cite{Gatteschi_Angew.Chem.Int.Ed.42/2003,Romeike2006May,Timm2006Jun,Jo2006Aug,Elste2007May,Misiorny2007Apr,Takahashi2009Feb,
Misiorny2009Apr,Parks2010Jun,Elste2010Jan,Andergassen2010Jun,Xie2012Mar,
Misiorny2013Jul,Misiorny2015Jan,Plominska2015Nov,Urdampilleta2015Apr,Plominska2017Apr,Plominska2018Jan}.
Due to the bistability of magnetic molecules \cite{Gatteschi_book,Bartolome}, such systems have
a great potential for applications in the information storing and processing technologies
\cite{Loss1998Jan,Ardavan2007Jan,Mannini2009Feb,Frankland2013Jan}.
Moreover, when attached to ferromagnetic contacts, magnetic molecules
can exhibit a large spin-valve effect when the magnetic configuration
of the device varies from the parallel to antiparallel alignment of leads' magnetic moments 
\cite{Barnas_J.Phys.:Condens.Matter20/2008,Misiorny2009Jun}.
Quite interestingly, recently, a spin-valve like behavior has also been found
in the case of nonmagnetic junctions involving single molecular magnets \cite{Urdampilleta2011Jul,Urdampilleta2011Oct}.
In particular, Urdampilleta et al. examined transport through a
supramolecular spin-valve---a tunnel junction with
an embedded carbon nanotube to which molecular magnets were attached.
By aligning the spins of magnetic molecules
with an external magnetic field, a change in the conductance of the 
system was observed. Such tuning of the current flowing through the system
without the necessity to use ferromagnetic contacts provides a perspective
way of the current control through the spin degrees of freedom,
which is important for molecular spintronics \cite{Rocha2005Apr,Bogani2008Mar,Awschalom2013Mar}.

Motivated by this experimental achievement, in this paper we theoretically investigate
the transport properties of a tunnel junction involving a single-wall carbon nanotube with 
two attached magnetic molecules in the presence of external magnetic field.
Our considerations are carried out by assuming a weak tunnel coupling between the nanotube
and external reservoirs, such that the current is mainly driven by sequential tunneling processes.
However, we also examine the role of cotunneling processes, which 
determine the magnetoresistive properties of the considered device in the low bias voltage regime.
By determining the currents flowing through the system in the absence
and presence of external magnetic field,
we analyze the behavior of the tunnel magnetoresistance of the device.
Our work quantifies thus the change of spin-resolved transport properties
when the state of the molecules switches from the superparamagnetic
to the ferromagnetic one.
Depending on the occupation of the carbon nanotube and the applied bias voltage,
we find transport regimes of both large positive magnetoresistive response
as well as transport regions where this response becomes negative.
The calculations are performed by using the real-time diagrammatic technique 
in the first and second order of expansion with respect to the tunnel coupling
\cite{Schoeller1994Dec,Konig1996Dec,Thielmann2005Sep}.

%
\section{Model and Hamiltonian}

\begin{figure}[t]
  \includegraphics[width=0.95\columnwidth]{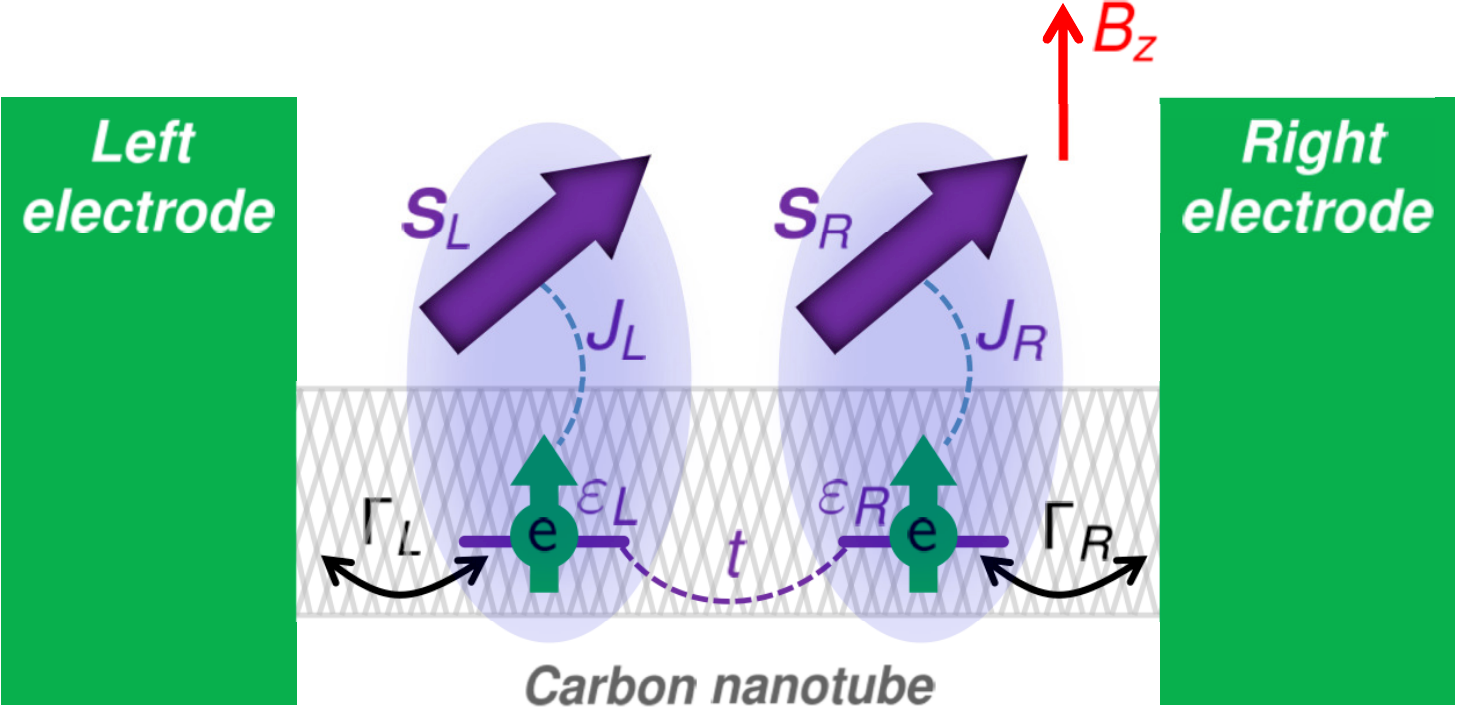}
  \centering
  \caption{\label{Fig:1}
  The schematic of a supramolecular spin valve. It consists of a single-wall carbon nanotube
  with attached two magnetic molecules. The nanotube
  is coupled to the left and right metallic leads.
  See the main text for details.}
  \label{fig1}
\end{figure}

The considered molecular system embedded between two nonmagnetic electrodes
is schematically shown in Fig.~\ref{fig1}.
It consists of two magnetic molecules of spin $S_{r}$
and magnetic anisotropy $D_{r}$, with $r=L$ ($r=R$)
for the left (right) molecule. The molecules
are exchange-coupled to the single-wall carbon nanotube with strength given by $J_r$.
It is assumed that the coupling of the molecules 
to the nanotube results in formation of orbital levels in nanotube in their vicinity,
through which transport takes place \cite{Urdampilleta2011Jul,Urdampilleta2011Oct}.
Thus, the total Hamiltonian of the system can be written as
\beq
H = H_{leads} + H_{mol} + H_{B} + H_{tun}.
\eeq
The first part of $H$ describes the noninteracting electrons
in the external electrodes and takes the form
\beq
  H_{leads} = \sum_{rk\s} \e_{rk} c^\dag_{rk\s} c_{rk\s},
\eeq
with $c^\dag_{rk\s}$ $(c_{rk\s})$ being the operator for creation (annihilation)
of a spin-$\s$ electron with momentum $k$ and the energy $\e_{rk}$ in the $r$th lead.
The second term of the total Hamiltonian characterizes the nanotube-magnetic molecule
subsystem and can be expressed as \cite{Timm2006Jun,Elste2007May,Misiorny2009Jun,Oreg2000Jul}
\begin{eqnarray}
 H_{mol} \!\!\!\!&=&\!\!\!\! \sum_r \e_{r} n_{r} + E_C(n \!-\! n_{0})^2
+ t\sum_\s(d^\dag_{L\sigma} d_{R\sigma} \!+\! h.c.)\nonumber\\
&&- \sum_r \left(J_{r} \mathbf{S}_{r} \! \cdot\! \mathbf{s}_{r} + D_{r} S_{rz}^2\right).
\end{eqnarray}
The particle number operator for an electron of spin $\s$ occupying the nanotube's orbital level $r$
is given by $n_{r\s} = d^\dag_{r\s}d_{r\s}$, and $n_r = n_{r\uparrow}+n_{r\downarrow}$,
$n = n_L + n_R$. The charge of a neutral nanotube is denoted by $n_0$.
$E_C$ is the charging energy of the nanotube and $t$ denotes the hopping between the two
orbital levels. The exchange coupling between molecule $r$
and nearby orbital level is denoted by $J_r$, with 
$\mathbf{S}_{r}$ and $\mathbf{s}_{r}$ denoting the operators for spin of the molecule
and the spin of electron on orbital level, respectively.
The magnetic anisotropy associated with the $r$th molecule is represented by $D_{r}$,
where $S_{rz}$ is the $z$th component of $\mathbf{S}_{r}$.
The third part of $H$ describes the external magnetic field
$B_{z}$ (in units of $g\mu_B\equiv 1$) applied to the molecular system
\be
    H_{B} = B_{z} S_{z}^{tot},
\ee
where $S_{z}^{tot} =  \sum_{r} (S_{rz} + s_{rz}$).
Finally, the last term of the total Hamiltonian 
describes the electron tunneling processes and can be written as
\be
    H_{tun} =  \sum_{r k\sigma} v_{r}[c^\dagger_{r k\sigma} d_{r \sigma} + d_{r \sigma}^\dag c_{r k\sigma}]
\ee
with $v_{r}$ denoting the respective tunnel matrix elements. 
The coupling between the $r$th electrode and $r$th orbital
level of the nanotube can be defined as $\G_{r} = 2\pi\rho_{r} v_{r}^2$,
with $\rho_{r}$ denoting the density of states of the $r$th lead at the Fermi level.
In our analysis, we take into account the symmetric case
assuming $\G_{r} \equiv \G$. Moreover, we also 
assume that the two molecules are identical, i.e. $S_r \equiv S$,
$D_r\equiv D$ and $J_r\equiv J$.


\section{Method}

In order to calculate the current flowing through the considered molecular device,
we use the real-time diagrammatic technique \cite{Schoeller1994Dec,Konig1996Dec,Thielmann2005Sep}
including the first and second order terms of expansion with respect to the tunnel coupling $\Gamma$.
In calculations, one first needs to determine the corresponding diagrams that
contribute to the elements $W_{\ket{\chi}\ket{\chi'}}$ of the self-energy matrix $\mathbf{W}$
in given order of expansion,
i.e. $\mathbf{W} = \mathbf{W}^{(1)} + \mathbf{W}^{(2)}$.
Here, $\ket{\chi}$ denotes the eigenstate of $H_{mol}$, $H_{mol}\ket{\chi} = \e_\chi \ket{\chi}$,
and $\e_\chi$ is the corresponding eigenenergy.
In the first-order of expansion the off-diagonal elements of $\mathbf{W}^{(1)}$ are given by
\begin{eqnarray*}
   W_{\chi \chi^\prime}^{(1)} &=& 2\pi \sum_{r\s } \rho_{r}
   \Big\{
   f_r (\varepsilon_\chi - \varepsilon_{\chi^\prime})
   \big|
   v_{r} \bra{\chi}d_{r\sigma}^\dagger
   \ket{\chi^\prime}\big|^2 \\
   &+& \left[ 1-f_r (\varepsilon_{\chi^\prime} - \varepsilon_{\chi}) \right]
   \big|v_{r} \bra{\chi}d_{r\sigma}
   \ket{\chi^\prime}\big|^2 \Big\},
\end{eqnarray*}
where $f_{r}(\varepsilon) = 1/[e^{(\varepsilon-\mu_{r})/T}+1]$ is the Fermi-Dirac distribution function
with the electrochemical potential of $r$th lead denoted as $\mu_{r}$
and $T$ standing for temperature ($k_{\rm B}\equiv 1$).
The diagonal elements of $\mathbf{W}^{(1)}$ are equal to
$W_{\chi \chi}^{(1)} = -\sum_{\chi^\prime\neq\chi} W_{\chi^\prime\chi}^{(1)}$.
The formulas for the second-oder self-energies ($\mathbf{W}^{(2)}$) are much more cumbersome
since they involve summations over many virtual states of the molecule \cite{Weymann2008Jul},
therefore, we will not present them here.
The self-energy matrices allow for the calculation of 
the corresponding probabilities of occupation of states $\ket{\chi}$
which can be done using the following equations \cite{Thielmann2005Sep}
\begin{equation}
    \mathbf{W}^{(1)}\mathbf{P}^{(0)}=0 \;\;\;\;
    {\rm and} \;\;\;\;
    \mathbf{W}^{(2)}\mathbf{P}^{(0)} + \mathbf{W}^{(1)}\mathbf{P}^{(1)}=0,
\end{equation}
where the vector of probabilities in given order is normalized such that 
${\rm Tr}\{\mathbf{P}^{(0)}\}=1$ and ${\rm Tr}\{\mathbf{P}^{(1)}\}=0$.
Then, the current in the first $(I^{(1)})$ and second order $(I^{(2)})$
of expansion can be found from \cite{Thielmann2005Sep}
\begin{eqnarray}
I^{(1)} &=& \frac{e}{2\hbar} {\rm Tr} \{ \textbf{W}^{I(1)}\textbf{P}^{(0)} \}, \\
I^{(2)} &=& \frac{e}{2\hbar} {\rm Tr} \{ \textbf{W}^{I(2)}\textbf{P}^{(0)} + \textbf{W}^{I(1)}\textbf{P}^{(1)} \}, 
\end{eqnarray}
respectively. Here, the self-energy matrices 
$\textbf{W}^{I(1)}$ and $\textbf{W}^{I(2)}$
are similar to $\textbf{W}^{(1)}$ and $\textbf{W}^{(2)}$ except for the
fact that they take into account the number of electrons transferred through the system.
The total current is then simply given by $I = I^{(1)} + I^{(2)}$.

In the following, we study the transport properties in the full parameter
space, i.e. in the full range of bias and gate voltages. However,
because the calculation of the second-order contribution to the current
in the full parameter space is a numerically demanding task,
we will discuss the role of the second-order processes only
in the low bias voltage regime, where such processes
play the most important role in transport \cite{Grabert1993}. For larger voltages,
sequential processes give a dominant contribution to the conductance and
therefore the transport properties of the system can be reliably
described including only first-order processes.
Therefore, we first discuss the transport behavior 
in the full parameter space considering sequential processes
and only later on we extend the discussion
to the case of cotunneling in the linear response regime.

\section{Results and discussion}

Our calculations are performed for the following parameters of the system.
Each magnetic molecule is characterized by spin $S=1$ and
the uniaxial magnetic anisotropy $D/E_C=0.2$.
The exchange coupling between the corresponding molecule and orbital level
in the nanotube is assumed to be of antiferromagnetic type 
\cite{Urdampilleta2011Jul,Urdampilleta2011Oct}, and we take $J/E_C = -0.4$.
The hopping between the orbital levels of the nanotube 
is assumed to be $t/E_C=0.2$, while the position of orbital levels
is characterized by the energy $\varepsilon$
with the assumption $\varepsilon_{L}=\varepsilon_{R}\equiv\varepsilon$.
The coupling to external leads is taken as $\Gamma/E_C=0.02$
and the calculations are performed at the temperature
$T/E_C = 0.08$. Finally, we use the charging energy as
the energy unit $E_C\equiv 1$.

\subsection{The differential conductance}

\begin{figure}[t!]
  \includegraphics[width=1\columnwidth]{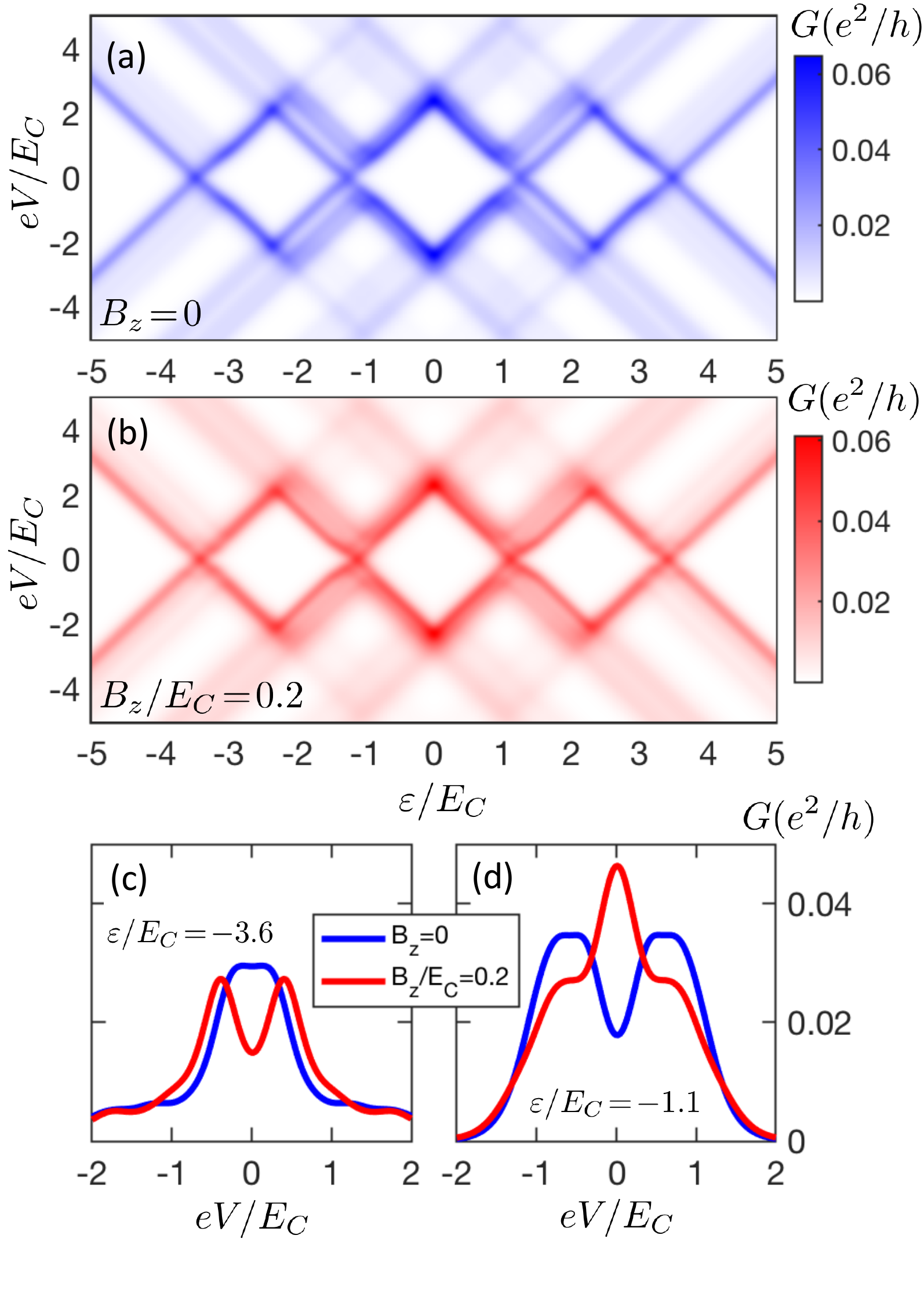}
  \caption{
  The differential conductance $G$ as a function of the bias voltage $V$
  and the energy of nanotube's orbital levels $\varepsilon$ in the case of (a) $B_z=0$ and (b) $B_z/E_C=0.2$.
  The bottom panels show the bias dependence of $G$
  calculated for (c) $\varepsilon/E_C=-3.6$ and (d) $\varepsilon/E_C=-1.1$.
  The parameters are:
  $S=1$, $D/E_C=0.2$, $J/E_C=-0.4$, $\Gamma/E_C=0.02$, $t/E_C=0.2$, $T/E_C=0.08$,
  and $E_C\equiv 1$.}
  \label{fig2}
\end{figure}

Let us start the discussion with the analysis of the behavior
of the differential conductance.
Figure~\ref{fig2} presents $G$ plotted as a function of the bias
voltage $V$ and the position of the nanotube's energy levels $\e$
in the case of (a) $B_{z}=0$ and (b) finite magnetic field.
In each case, one can observe typical Coulomb diamond patterns
associated with single-electron charging effects.
With lowering $\e$, the nanotube becomes occupied
by electrons and each time the two charge states
become degenerate there is a resonance in the linear response conductance.
In-between the resonances the molecule is in the Coulomb blockade regime,
where sequential tunneling processes are exponentially suppressed.
The electrons start to tunnel through the molecular system
when the applied bias voltage exceeds some threshold.
Then, there appears a step in the current and associated
differential conductance peak. For larger voltages,
excited states start playing a role resulting
in additional peaks in the conductance.
This behavior is clearly associated with discrete
energy spectrum of the nanotube-molecule system and,
thus, is present irrespective of the value of magnetic field.
The presence of $B_z$, however,  spin-slits 
the levels and therefore changes the energies of 
molecular states. As a consequence, the conductance spectra
become slightly modified, cf. Figs. \ref{fig2}(a) and (b).

To exemplify the supramolecular spin valve effect, in
Figs.~\ref{fig2}(c) and (d) we present the bias voltage dependence
of the differential conductance for two different values of $\e$,
as indicated. When $B_z=0$, the molecules are in a superparamagnetic state
and their spins become aligned only when external magnetic
field is applied to the system. Thus, by manipulating the spins
of the molecules, it is possible to change the current flowing through
the whole device. However, the behavior of the system is 
not as simple as one could expect, i.e. depending on $\e$
and $V$, one can find both regimes of the current suppression
or its enhancement with the application of magnetic field.
Such behavior can be seen in Figs. \ref{fig2}(c) and (d).

\subsection{The tunnel magnetoresistance}

\begin{figure}[t!]
  \includegraphics[width=0.9\columnwidth]{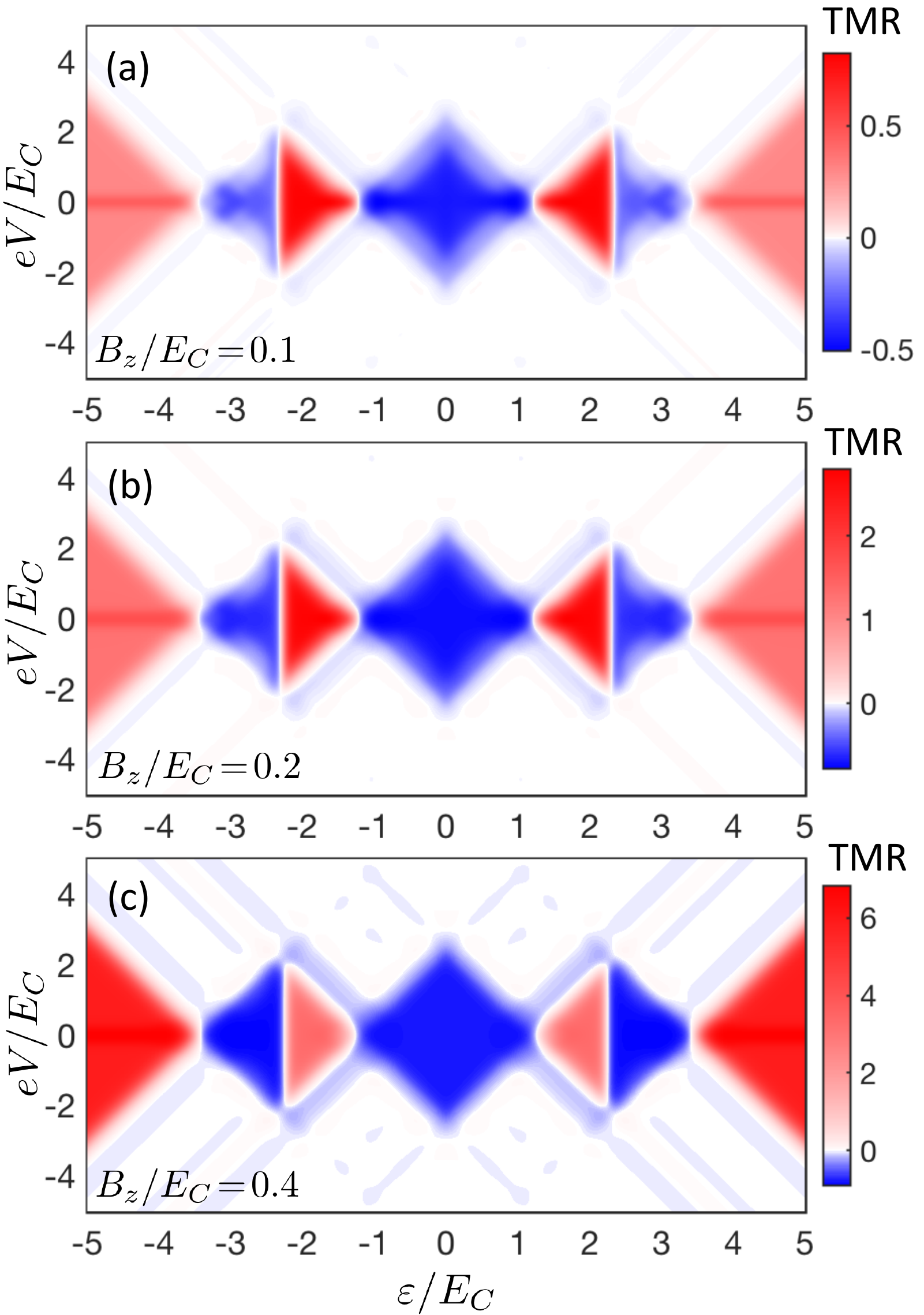}
  \centering
  \caption{
  The bias voltage and nanotube's orbital level
  dependence of the tunnel magnetoresistance (TMR)
  for a few values of external magnetic field, as indicated.
  The other parameters are the same as in Fig.~\ref{fig2}.}
  \label{fig3}
\end{figure}

To quantify the change of the system's transport properties
in the presence and absence of external magnetic field,
we introduce the tunnel magnetoresistance (TMR), defined as 
\begin{equation}
{\rm TMR} = \frac{I(B_z=0)-I(B_z)}{I(B_z)},
\end{equation}
where $I (B_z)$ is the current flowing through the system
in the presence of magnetic field $B_z$.

The bias voltage and level position dependence of the TMR
for a few values of magnetic filed is presented in \fig{fig3}.
First of all, one can note that the TMR is very low in the
large bias voltage regime. In this case many molecular states
of the system are relevant for transport and 
sequential tunneling dominates the current.
Consequently, the difference between
the currents flowing in the case when $B_z=0$
and $B_z\neq 0$ is very small, which results in ${\rm TMR}\approx 0$,
see \fig{fig3}. The situation becomes however completely different
when only a few molecular states contribute to the current,
which happens in the low-bias voltage regime.
We note that in this transport regime
the current can flow due to thermally-activated
sequential tunneling processes, which
in the case of $\Gamma \ll T$ are relevant,
as well as cotunneling processes.
In \fig{fig3} we show the results due to the first-order 
processes, while the role of cotunneling
will be analyzed further on.

Generally, one can see that for low bias voltages
the TMR strongly depends on the
charge state of the nanotube. More specifically, in the two-electron
Coulomb diamond (around $\e = 0$, see \fig{fig3}), the TMR
is negative, while in the case when the nanotube is either 
fully occupied or empty
($|\e/E_C| \gtrsim 3.5$), the TMR is positive.
On the other hand, in the Coulomb blockade regimes
with odd number of electrons on the nanotube, the TMR can be both
negative and positive, depending on $\e$, see \fig{fig3}
for $1\lesssim |\e/E_C| \lesssim 3.5$.
To understand this behavior, one needs to consider the corresponding molecular
states relevant for transport in each region.
When the nanotube is either empty or fully occupied,
the presence of attached magnetic molecules
is not that important for low bias voltages. Because with increasing
the value of magnetic field, the number of 
states relevant for thermally-activated transport decreases
due to the Zeeman splitting,
$I(B_z) < I(B_z=0)$, and consequently ${\rm TMR}>0$.

This is however opposite to the case when the nanotube is occupied
by two electrons, where one finds $I(B_z) > I(B_z=0)$, yielding ${\rm TMR}<0$.
The reason for this behavior is associated with the fact
that in the presence of magnetic field the spins
of the molecules become aligned, such that the
molecule-nanotube system is mainly occupied
by a two-electron state with $S_z^{tot} = -2S$
being a linear combination of local states with
one electron on each level and zero and two electrons on different levels.
This increases transport compared to the case of no magnetic field,
where the occupation probability is distributed between several two-electron states.

When the nanotube is occupied by an odd number of electrons
at low voltages the system is in the state with $S_z^{tot} = -2S+1/2$.
In this case it is relevant whether the excitation energies to charge
states with empty (fully occupied) nanotube and states with two electrons on the nanotube
are more favorable. In the former case the current becomes suppressed
in finite $B_z$, whereas in the latter case the current
gets enhanced with the application of magnetic field, see \fig{fig3}.
It is also interesting to note that the above-described behavior
becomes enhanced with increasing $B_z$,
resulting in larger $|{\rm TMR}|$, cf. \fig{fig3}.

\subsection{Current spin polarization}

\begin{figure}[t!]
  \includegraphics[width=0.9\columnwidth]{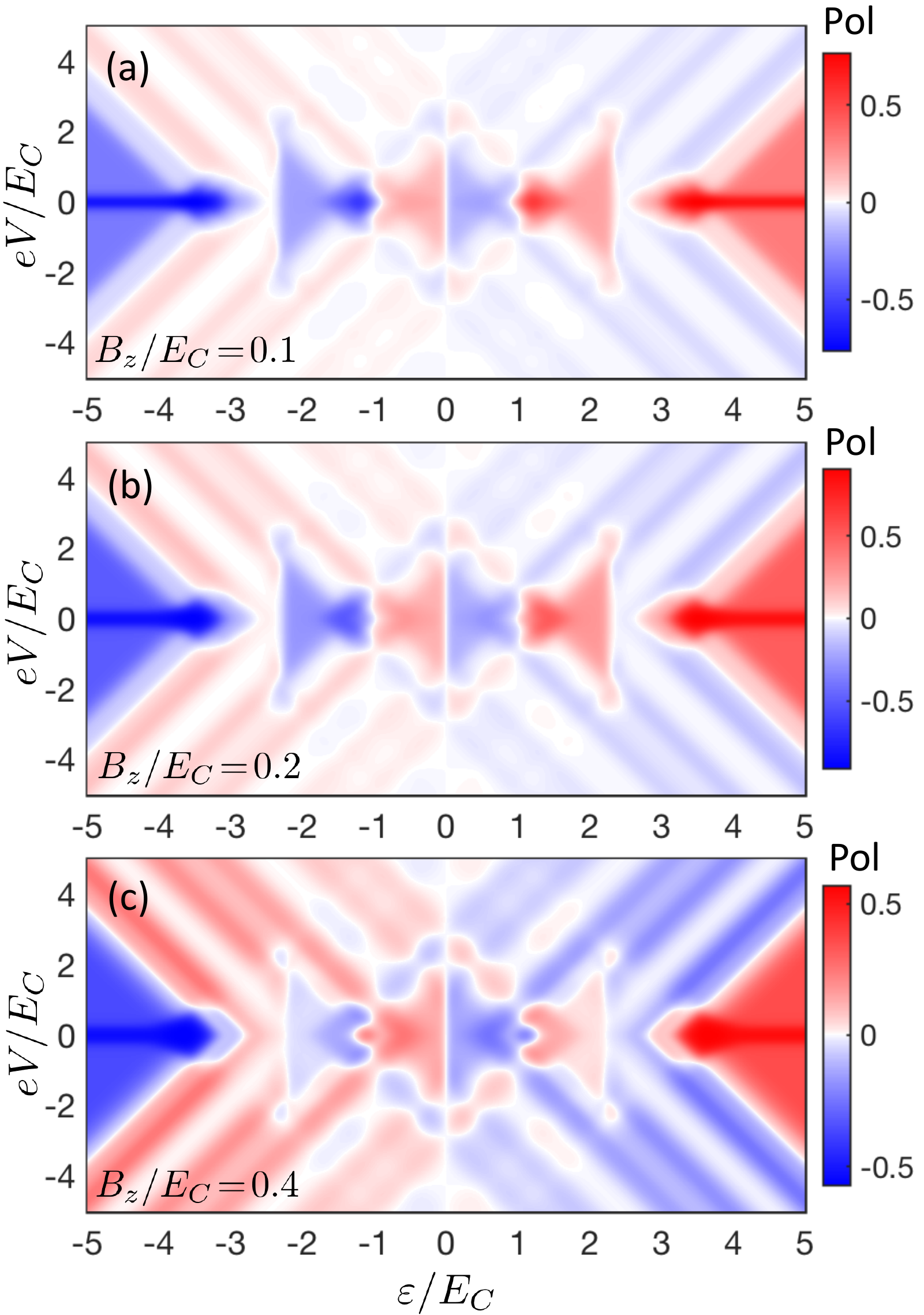}
  \centering
  \caption{
  The dependence of the current spin polarization (Pol) on 
  bias voltage $V$
  and the nanotube's orbital level position $\varepsilon$
  for a few different values of external magnetic field, as indicated.
  The parameters are the same as in Fig.~\ref{fig2}.}
  \label{fig4}
\end{figure}

Let us now analyze how the spin polarization of the flowing current
changes with increasing external magnetic field. The spin polarization is defined as
\begin{equation}
{\rm Pol} = \frac{I_\uparrow(B_z)-I_\downarrow(B_z)}{I_\uparrow(B_z)+I_\downarrow(B_z)},
\end{equation}
where $I_\sigma(B_z)$ is the current flowing in the spin-channel $\sigma$
at magnetic field $B_z$. The spin polarization as a function
of the bias voltage $V$ and the nanotube's level position $\e$ for different values of $B_z$ is shown in 
\fig{fig4}. First of all, one can note that the dependence is symmetric
with respect to the particle-hole symmetry point of the model,
i.e. $\e=0$, with ${\rm Pol}(\e<0) = -{\rm Pol}(\e>0)$.
Moreover, as in the case of TMR discussed above
one observed a gradual increase
of $|{\rm TMR}|$ with boosting $B_z$, now this is not the case.
More specifically, $|{\rm Pol}|$ increases when $B_z$ grows from 
$B_z/E_C=0.05$ to $B_z/E_C=0.1$, however, then it slightly drops
when magnetic field is raised further to $B_z/E_C=0.2$, see \fig{fig4}.
A general observation is that out of the Coulomb blockade regime,
${\rm Pol}<0$ (${\rm Pol}>0$) for $\e>0$ ($\e<0$).
On the other hand, the largest magnitude of the spin polarization is found 
in the low bias voltage regime when the nanotube is either empty or fully occupied.
One then finds ${\rm Pol}>0$ (${\rm Pol}<0$) for $\e/E_C \gtrsim 3.5$ ($\e/E_C \lesssim -3.5$),
which is simply associated with the fact that for $\e/E_C \gtrsim 3.5$ ($\e/E_C \lesssim 3.5$)
the excitations to positive (negative) spin states are more favorable.

\subsection{Effects of cotunneling}

\begin{figure}[t!]
  \includegraphics[width=1\columnwidth]{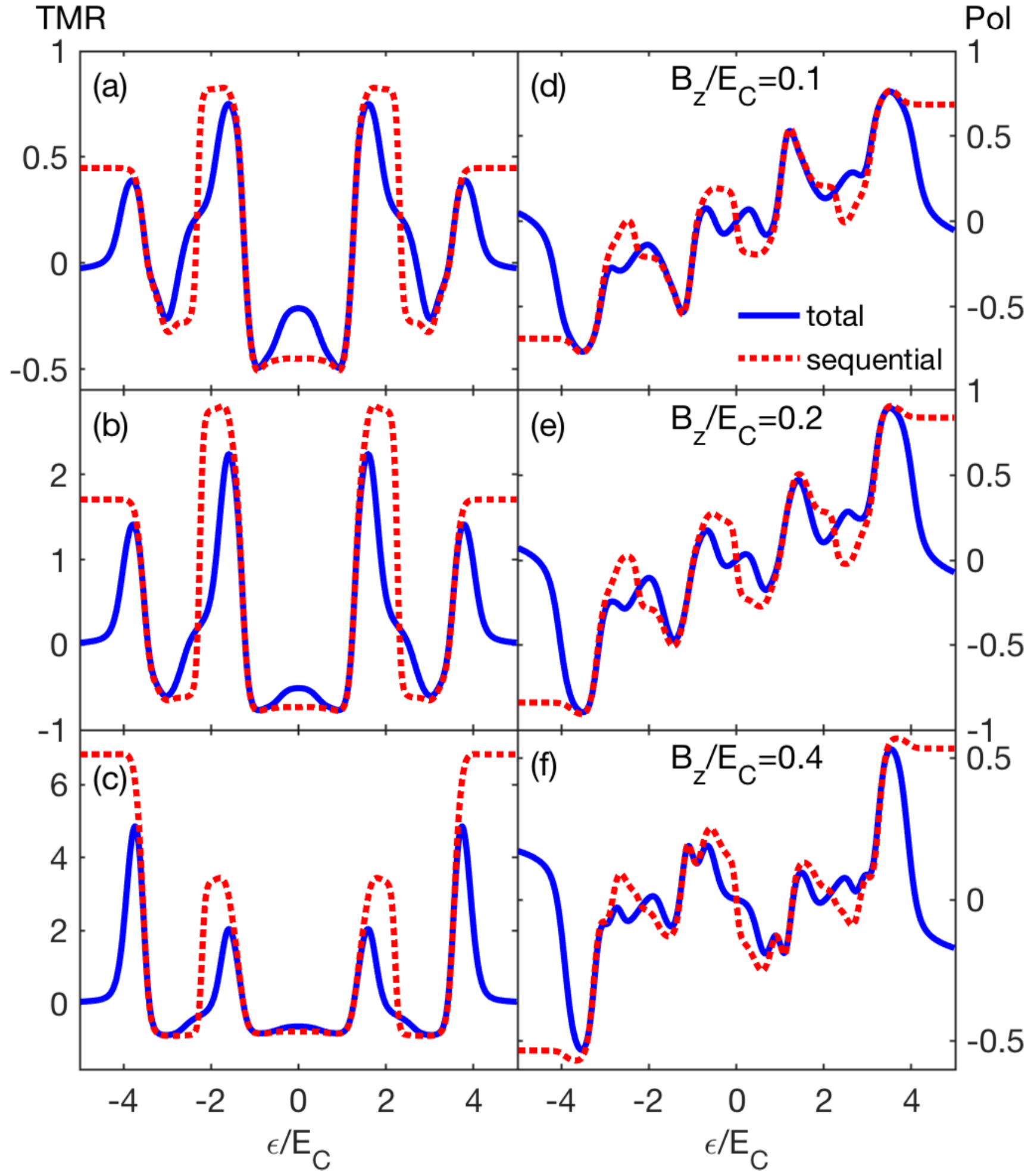}
  \caption{
  The dependence of the tunnel magnetoresistance (left column) and
  spin polarization (right column) on the position
  of the nanotube's orbital levels $\varepsilon$,
  for selected values of magnetic field,
  calculated in the linear response regime.
  Solid lines present the results
  obtained using the first and second order processes,
  while the dotted lines show the sequential tunneling results.
  The parameters are the same as in Fig.~\ref{fig2}.}
  \label{fig5}
\end{figure}

Finally, in this section we discuss the role of cotunneling
processes on the tunnel magnetoresistance and
spin polarization of the flowing current.
Figure \ref{fig5} shows the total (sequential plus cotunneling) 
TMR as well as the total spin polarization
as a function of the nanotube's level position $\e$
calculated in the linear response regime.
For comparison, we also show the results
obtained by considering only the first-order tunneling processes.
One can easily see the differences between both results,
which are most revealed in the case of empty or fully 
occupied nanotube, $|\e/E_C| \gtrsim 3.5$, where elastic non-spin-flip 
cotunneling processes are most important for the current.
There, one finds a strong suppression
of the tunnel magnetoresistance, such that ${\rm TMR} \approx 0$,
which is completely opposite to the sequential tunneling result,
see the left column of \fig{fig5}.
Because in this transpose regime elastic cotunneling processes,
in which the spin of tunneling electrons is conserved,
drive the current, the difference between the currents
$I(B_z=0)$ and $I(B_z)$ decreases as one moves deeper and deeper
into the empty or fully-occupied nanotube regime, cf. Figs. \ref{fig5}(a)-(c).
A similar observation also applies to the behavior of the current spin polarization,
which becomes generally suppressed compared to that 
predicted by considering only sequential tunneling processes,
see the right column of \fig{fig5}.
On the other hand, as far as other transport regimes with different nanotube occupations are concerned,
it can be concluded from \fig{fig5} that sequential tunneling processes
give a qualitatively reliable insight into
the transport behavior of the system. In these transport regimes the cotunneling
processes rather weakly modify the observed behavior.

\section{Summary}

We have analyzed the magnetoresistive properties of
a supramolecular spin valve consisting of a nanotube with two attached magnetic molecules
embedded in a tunnel junction. The considerations were
performed by using the real-time diagrammatic technique in 
the first and second-order of expansion with respect to the tunnel coupling.
We have shown that the tunnel magnetoresistance 
of such device, associated with a change of 
magnetic molecules' state from the superparamagnetic to the ferromagnetic one,
can take both positive and negative values, depending on the transport regime.
Our work demonstrates thus that it is possible
to tune the TMR by either the bias or gate voltage.
This offers an interesting route for the control of the magnetoresistive transport properties
without the need to use ferromagnetic contacts.
In addition, we have also studied the spin polarization
of the tunneling current and shown that it strongly
depends on the transport regime, which allow for tuning
both the magnitude and sign of the spin polarization.


\begin{acknowledgements}
We acknowledge discussions with J\'ozef Barna\'s.
This work was supported by the National Science Centre
in Poland as the Project No. DEC-2013/10/E/ST3/00213.
Computing time at Pozna\'n Supercomputing and Networking Center is acknowledged.
\end{acknowledgements}


\bibliographystyle{eplbib}
\bibliography{Bibliography}

\end{document}